# Balancing Between Over-Weighting and Under-Weighting in Supervised Term Weighting

Haibing Wu[1], Xiaodong Gu[2]
Department of Electronic Engineering, Fudan University
[1]haibingwu13@fudan.edu.cn, [2]xdgu@fudan.edu.cn

**Abstract**

Supervised term weighting could improve the performance of text categorization. A way proven to be effective is to give more weight to terms with more imbalanced distributions across categories. This paper shows that supervised term weighting should not just assign large weights to imbalanced terms, but should also control the trade-off between over-weighting and under-weighting. Over-weighting, a new concept proposed in this paper, is caused by the improper handling of singular terms and too large ratios between term weights. To prevent over-weighting, we present three regularization techniques: add-one smoothing, sublinear scaling and bias term. Add-one smoothing is used to handle singular terms. Sublinear scaling and bias term shrink the ratios between term weights. However, if sublinear functions scale down term weights too much, or the bias term is too large, under-weighting would occur and harm the performance. It is therefore critical to balance between over-weighting and under-weighting. Inspired by this insight, we also propose a new supervised term weighting scheme, regularized entropy (*re*). Our *re* employs entropy to measure term distribution, and introduces the bias term to control over-weighting and under-weighting. Empirical evaluations on topical and sentiment classification datasets indicate that sublinear scaling and bias term greatly influence the performance of supervised term weighting, and our *re* enjoys the best results in comparison with existing schemes.

## 1   Introduction

Baseline approaches to text categorization often involve training a linear classifier over bag-of-term representations of documents. In such representations, a textual document is represented as a vector of terms. Terms can be words, phrases, or other more complicated units identifying the contents of a document. Given that some terms are more informative than others, a common technique is to apply a term weighting scheme to give more weight to discriminative terms and less weight to non-discriminative ones. Term weighting schemes fall into two categories. The first one, known as unsupervised term weighting, does not take category information into account. Inverse document frequency (*idf*) is a commonly used unsupervised method. The second one referred to as supervised term weighting embraces the category label information of training documents in the categorization tasks (Batal, & Hauskrecht, 2009; Soucy, & Mineau, 2005; Debole & Sebastiani, 2003; Wu, & Gu, 2014). Many supervised term weighting schemes have been studied in the literature and show better or worse results than standard *idf* (see section 2 for details). So a natural question is: what is the key to a successful supervised term weighting scheme?

This paper advocates that supervised term weighting should (1) assign larger weights to terms with more imbalanced distributions across categories, and (2) balance between over-weighting and under-weighting, both of which are caused by unsuitable quantification of term's distribution. Over-weighting, a new proposed concept, would occur due to the improper handling of singular terms and unreasonably too large ratios between term weights. To reduce over-weighting, we present three regularization techniques: add-one smoothing, sublinear scaling and bias term. Singular terms feature high imbalanced distributions across categories. Singular terms could be very discriminative, or noisy and useless. Add-one smoothing, a commonly used technique, is introduced to handle singular terms. Sublinear scaling and bias term shrink the ratios between term weights, and thus prevent over-weighting. However, if the sublinear functions scale down term weights too much, or the bias term is too large, ratios between term weights will become too small, leading to under-weighting problem. So a well-performed supervised term weighting scheme should control the trade-off between over-weighting and under-weighting.

Inspired by the insight of giving large weights to imbalanced terms and controlling the trade-off between over-weighting and under-weighting, we also propose a novel supervised term weighting scheme, regularized entropy (*re*). *re* bases on entropy to measure the imbalance of term's distributions across categories, and assigns larger weights to terms with smaller entropy. The bias term in *re* controls over-weighting and under-weighting. If its value is too small, over-weighting occurs; conversely if too large, under-weighting occurs.

After presenting over-weighting, regularization and the *re* scheme, experiments are conducted on both topical and sentiment classification tasks. In our experiments, we first compare *re* with many existing term weighting schemes. The experimental results show that *re* performs well on all datasets, including sentiment and topical, balanced and imbalanced datasets. Specially, it achieves the best results on 9 of 14 tasks. Then we experimentally demonstrate that scaling functions greatly influence the performance of supervised term weighting. Additionally, we empirically analyse the effect of bias term in *re*, showing that the performance of *re* and the value of bias term exhibits an inverted U-shaped relationship.

## 2  Review of Term Weighting Schemes

One of the main issues in text categorization is the representation of documents. Vector Space Model (VSM) provides a simplifying representation by representing documents as vector of terms. Term weighting aims to evaluate the relative importance of different terms. There are three components in a term weighting scheme: local weight, global weight and normalization factor (Salton, & Buckley, 1988; Lan et al., 2009). Final term weight is the product of the three components:

$$x_{ij} = l_{ij} \times g_i \times n_j \tag{1}$$

Here $x_{ij}$ is the final weight of $i_{th}$ term in the $j_{th}$ document, $l_{ij}$ is the local weight of $i_{th}$ term in the $j_{th}$ document, $g_i$ is the global weight of the $i_{th}$ term, and $n_j$ is the normalization factor for the $j_{th}$ document.

| # | Local weight | Formulation | Description |
|---|---|---|---|
| 1 | *tf* | $tf$ | Raw term frequency. |
| 2 | *tp* | $\begin{cases} 1, & \text{if } tf > 0 \\ 0, & \text{otherwise} \end{cases}$ | Term presence, 1 for presence and 0 for absence. |
| 3 | *atf* | $k + (1-k)\dfrac{tf}{\max_t(tf)}$ | Augmented term frequency, $\max_t(tf)$ is the maximum frequency of any term in the document, $k$ is set to 0.5 for short documents (Salton, & Buckley, 1988; Croft, 1983). |
| 4 | *ltf* | $\log_2(1+tf)$ | Logarithm of term frequency. |
| 5 | *btf* | $\dfrac{(k_1+1)tf}{k_1\left((1-b)+b\dfrac{dl}{avg\_dl}\right)+tf}$ | BM25 *tf*, *aver_dl* is the average number of terms in all the documents. Default $k_1$ and $b$ parameters of BM25 are 1.2 and 0.95 respectively (Jones et al. 2000). |

Table 1. Local term weighting schemes.

### 2.1  Local term weighting

Local weight is derived only from frequencies within the document. Table 1 presents five common local weighting schemes: raw term frequency (*tf*), term presence (*tp*), logarithm of term frequency (*ltf*), augmented term frequency (*atf*) and BM25 term frequency (*btf*). The most popular and notable representation, *tf*, counts how many times the term occurs in a document. This means that *tf* gives more confidence to words that appears more frequently. The simplest binary representation, *tp*, ignores the occurrences of the term in the document. This can be useful when the number of times a word appears is not considered important.

The *atf* scheme is a combination of *tp* and *tf*. It tries to give confidence to any term that appears and then give some additional confidence to terms that appear frequently. Logarithmic function in *ltf* is used to adjust within-document frequency because a term that appears ten times in a document is not necessarily ten times as important as a term that appears once in that document. The most sophisticated local weighting method in table 1 is *btf*. It bases on a probabilistic model for IR.

| Notation | Description |
|---|---|
| $a$ | Positive document frequency, i.e., number of training documents in the positive category containing term $t_i$. |
| $b$ | Number of training documents in the positive category which do not contain term $t_i$. |
| $c$ | Negative document frequency, i.e., number of training documents in the negative category containing term $t_i$. |
| $d$ | Number of training documents in the negative category which do not contain term $t_i$. |
| $N$ | Total number of documents in the training document collection, $N = a + b + c + d$. |
| $N^+, N^-$ | $N^+$ is number of training documents in the positive category, and $N^-$ is number of training documents in the negative category. $N^+ = a + b$, $N^- = c + d$. |

Table 2. Notations used to formulate global term weighting schemes.

| # | Global weight | Formulation | Description |
|---|---|---|---|
| 1 | *idf* | $\log_2 \frac{N}{a+c}$ | Inverse document frequency (Jones, 1972) |
| 2 | *pidf* | $\log_2 \left( \frac{N}{a+c} - 1 \right)$ | Probabilistic *idf* (Wu, & Salton, 1981) |
| 3 | *bidf* | $\log_2 \frac{b+d+0.5}{a+c+0.5}$ | BM 25 *idf* (Jones et al., 2000) |
| 4 | *ig* | $\frac{a}{N} \log_2 \frac{aN}{(a+b)(a+c)} + \frac{b}{N} \log_2 \frac{bN}{(a+b)(b+d)} + \frac{c}{N} \log_2 \frac{cN}{(a+c)(c+d)} + \frac{d}{N} \log_2 \frac{dN}{(b+d)(c+d)}$ | Information gain |
| 5 | *gr* | $\frac{ig}{-\frac{a+b}{N} \log_2 \frac{a+b}{N} - \frac{c+d}{N} \log_2 \frac{c+d}{N}}$ | Gain ratio |
| 6 | *mi* | $\log_2 \left( \max(\frac{aN}{(a+c)N^+}, \frac{cN}{(a+c)N^-}) \right)$ | Mutual information |
| 7 | *chi* | $\frac{N(ad-bc)^2}{(a+c)(b+d)(a+b)(c+d)}$ | Chi-square |
| 8 | *didf* | $\log_2 \frac{N^- a}{N^+ c}$ | Delta *idf* (Martineau, & Finin, 2009) |
| 9 | *dsidf* | $\log_2 \frac{N^-(a+0.5)}{N^+(c+0.5)}$ | Delta smoothed *idf* (Paltoglou, & Thelwall, 2010) |
| 10 | *dspidf* | $\log_2 \frac{(N^- - c)(a+0.5)}{(N^+ - a)(c+0.5)}$ | Delta smoothed *idf* (Paltoglou, & Thelwall, 2010) |
| 11 | *dbidf* | $\log_2 \frac{(N^- - c + 0.5)(a+0.5)}{(N^+ - a + 0.5)(c+0.5)}$ | Delta BM25 *idf* (Paltoglou, & Thelwall, 2010) |
| 12 | *rf* | $\log_2 \left( 2 + \frac{a}{\max(1,c)} \right)$ | Relevance frequency (Lan et al., 2009) |

Table 3. Global term weighting schemes.

## 2.2 Global term weighting

Global weight depends on the whole training document collection. Global term weighting tries to give a discrimination value to each term and place emphasis on terms that are discriminating. To formulate different global weighting schemes, some notations are first introduced in table 2. With these notations, table 3 presents several representative global term weighting schemes.

Scheme 1-3 in table 3, *idf*, *pidf* and *bidf* are unsupervised as they do not utilize the category label information of training documents. The *idf* is defined as the logarithm of the ratio of number of documents in a collection to the number of documents containing a specific term. Another two methods, *pidf* and *bidf*, are the variants of *idf*. The common idea behind *idf*, *pidf* and *bidf* is that a term occurring rarely is good at discriminating between documents. This idea is effective in IR, but not reasonable for text categorization. Because the text categorization task aims to discriminate between different categories, not documents.

Scheme 4-12 in table 3 are supervised term weighting schemes. Among these methods, *ig*, *gr*, *mi* and *chi*, are widely used for feature selection. Since feature selection selects relevant and valuable terms for text categorization, many feature selection methods are explored for term weighting in the literature. Debole and Sebastiani (2003) replaced *idf* with *ig*, *gr*, and *chi* for global term weighting. Experiments on Reuters-21578 showed that these feature selection methods did not give a consistent superiority over the standard *idf*. Batal and Hauskrecht (2009) empirically showed that *ig* and *chi* could greatly boost the performance of KNN classifier. Deng et al. (2014) also employed several feature selection methods, including *ig*, *mi* and *chi*, to learn the global weight of each term from training documents. Experimental results showed that compared with *bidf*, *mi* and *chi* produced better accuracy on two of three datasets, but *ig* performed very poorly.

Another supervised scheme is relevance frequency (*rf*) (Lan et al., 2009). The intuitive consideration of *rf* is that the more concentrated a high frequency term is in the positive category than in the negative category, the more contributions it makes in selecting the positive samples from the negative samples. Driven by this intuition, *rf* was proposed to capture this basic idea. Due to the asymmetry, *rf* only boosts the weights of terms that appear more frequently in the positive category. For terms appear more frequently in the negative category, it does not boost but decreases the weights of these terms. In other words, *rf* discriminates against terms appearing more frequently in negative category.

In sentiment analysis literature, Martineau and Finin (2009) presented a new supervised scheme *didf*. It is the difference of a term's *idf* values in the positive and negative training documents. Clearly, *didf* assigns large weights to terms with unevenly distribution between the positive and negative categories and discounts evenly distributed terms. A problem with *didf* is susceptible to the errors caused by the case of $a = 0$ or $c = 0$. Following the idea of *didf* and to rectify the problem of *didf*, Paltoglou and Thelwall (2010) presented a smoothed version of *didf*, delta smoothed *idf* (*dsidf*).[1] They also explored other more sophisticated methods originated from IR such as delta smoothed probabilistic *idf* (*dspidf*) and delta BM25 *idf* (*dbidf*). Empirical evaluations revealed that these smoothed delta variants of the classic *idf* scheme provided significant increase over the best term weighting methods for sentiment analysis in terms of accuracy.

## 2.3 Normalization

Normalization eliminates the document length effect. A common method is cosine normalization. Suppose that $t_{ij}$ represents weight of $i_{th}$ term in the $j_{th}$ document, then the cosine normalization factor is defined as $1/\sqrt{\sum_i t_{ij}^2}$.

## 3 Methodology

Our review of term weighting schemes above shows that supervised term weighting can, but not always, boost the performance of text categorization. Actually, the somewhat successful ones, such as *rf* and *didf*, follow the intuition that the more imbalanced a term's distribution across different categories, the more contribution it makes in discriminating between the positive and negative documents. The difference between them is the quantification of the degree of the imbalance of term's distributions. We argue that a successful supervised term weighting scheme should not just assign large weights to terms with imbalanced distributions, but should also balance between over-weighting and under-weighting when quantifying term's distributions across different categories.

---

[1] The formulation of *dsidf* in Paltoglou and Thelwall (2010) is *dsidf* = $\log_2(N^- a + 0.5 / N^+ c + 0.5)$, which suffers from over-weighting severely (see section 3.1 for details). So in table 3, we formulate *dsidf* as $\log_2 N^-(a+0.5)/N^+(c+0.5)$.

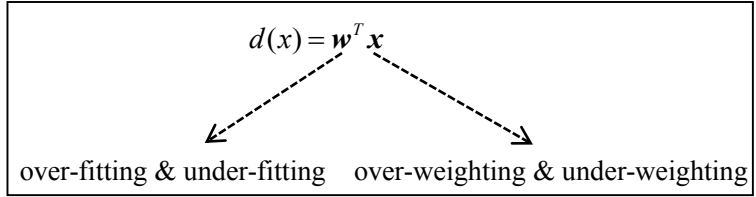

Figure 1. Over-fitting & under-fitting and over-weighting & under-weighting. Over-fitting and under-fitting occur with model parameter *w*. Over-weighting and under-weighting occur with term weight *x*. $d(x) = w^T x$ is the decision function for a text classifier.

### 3.1 Over-weighting & under-weighting and regularization

Over-weighting (and under-weighting) in supervised term weighting is somewhat like over-fitting (and under-fitting) in statistical machine learning. Let $d(x) = w^T x$ be the decision function for a linear text classifier. Over-fitting and under-fitting occur with model parameter *w*. Over-fitting arises when a statistical model describes random error or noise instead of the underlying relationship. Similarly, over-weighting and under-weighting occur with term weight *x* (see figure 1). In practice we identify that over-weighting is caused by the improper handling of singular terms and too large ratios between term weights.

The improper handling of singular terms would lead to the problem of over-weighting. Here we define singular terms as terms with high imbalanced distributions. Singular terms with high frequency could be very discriminative, deserving large term weights. But low frequency singular terms could be noisy and useless, and should be assigned small weights. Improper handling of singular terms could result in over-weighting. To illustrate this, suppose the training document collection is balanced with $N^+ = N^- = 1000$, term $t_1$ with $a = 100$ and $c = 0$, and term $t_2$ with $a = 2$ and $c = 0$. For comparison, we also include $t_3$ with $a = 100$ and $c = 1$. According to some existing schemes such as $dsidf = \log_2(N^- a + 0.5)/(N^- c + 0.5)$ in (Paltoglou, & Thelwall, 2010), the global weight of $t_1$, $t_2$ and $t_3$ is $g_1 = 17.6$, $g_2 = 12.0$ and $g_3 = 6.6$ respectively. This result violates our intuition that the weight of $t_1$ and $t_3$ should be large, and the weight of $t_2$ should be relatively small. Since the document frequency of $t_2$ is so trivial compared to the size of training collection, $t_2$ could be an unusual or noisy word. Another unreasonable observation is that the weight of $t_1$ seems too large compared to $t_3$, as both their frequency and distribution are close to each other. The unsuitable implementation of add-one smoothing[2] in this *dsidf* leads to unreasonably too large weight for both high and low frequency singular terms. Using the add-one smoothing as $\log_2 N^-(a+0.5)/N^-(c+0.5)$, then $g_1 = 7.7$, $g_2 = 1.3$ and $g_3 = 6.1$. This result satisfies our intuition that the weights of $t_1$ and $t_3$ should be close and large, and the weight of $t_2$ should be small.

An unsuitable quantification of term distribution may lead to unreasonably too large ratios between term weights and thus results in over-weighting. Let *x* be a variable quantifying term's distribution across categories. Here we define *x* as $x = \max(r^+, r^-)/\min(r^+, r^-)$, where $r^+ = (a+1)/N^+$, $r^- = (c+1)/N^-$. Obviously, *x* itself quantifies term distribution and could be used as global term weight. However, directly using *x* as global term weight often leads to unreasonably too large ratios between term weights. Because a term with $x = 100$ is not necessarily ten times as discriminative as a term with $x = 10$. To change the ratios between term weights, an intuitive regularization technique is to introduce sublinear scaling functions of *x*. This paper proposes and compares seven different scaling functions: $f1(x) = x^2$, $f2(x) = x^{1/2}$, $f3(x) = x^{1/3}$, $f4(x) = \log_2 x$, $f5(x) = 1/(0.1 + 1/x)$, $f6(x) = 1/(0.05 + 1/x)$ and $f7(x) = x^{1/6}$. For comparison, we also include $f0(x) = x$, which does not change the ratios between term weights. These functions are illustrated in figure 2. In these functions, $f1(x)$ dramatically amplifies ratios between term weights, increasing the risk of over-weighting. Conversely, $f7(x)$ may shrink the ratios too much, making no difference between discriminative and non-discriminative terms, hence leading to under-weighting.

---

[2] The "one" of add-one smoothing for $\log_2(N^- a + 0.5)/(N^- c + 0.5)$ is actually 0.5.

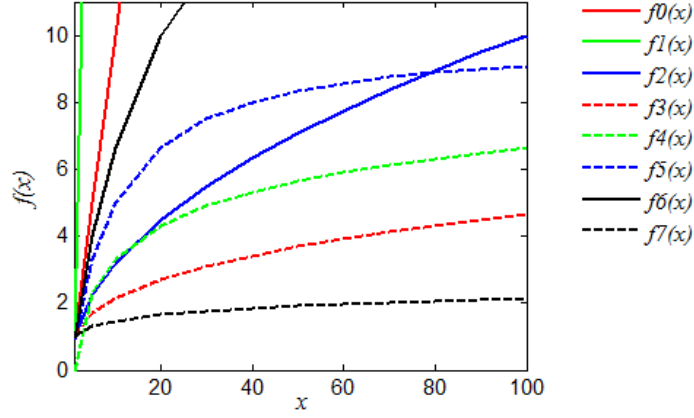

Figure 2. Illustration of different scaling functions: $f0(x)=x$, $f1(x)=x^2$, $f2(x)=x^{1/2}$, $f3(x)=x^{1/3}$, $f4(x)=\log_2 x$, $f5(x)=1/(0.1+1/x)$ $f6(x)=1/(0.05+1/x)$ and $f7(x)=x^{1/6}$.

In addition to sublinear scaling, we also propose another regularization technique, bias term, to further shrink the ratios between term weights. With bias term, the global term weight is formulated as follows:

$$g_i = b_0 + (1-b_0)f(x). \qquad (2)$$

Here $f(x)$, which should be normalized to [0,1], is a measurement quantifying term's distribution across categories, and $b_0 \in [0,1]$ is the bias term. The value of $b_0$ controls the trade-off between over-weighting and under-weighting. If $b_0$ is too large, under-weighting would occur. If $b_0$ is too small, over-weighting would occur. The optimal value of $b_0$ can be obtained via cross-validation, a model selection technique widely used in machine learning.

### 3.2 Regularized entropy

Inspired by the insight of balancing between over-weighting and under-weighting, we propose a novel supervised term weighting scheme, regularized entropy (*re*). For *re*, entropy is exploited to measure the degree of the imbalance of a term's distribution across different categories. According to information theory (Shannon, 1948), for a random variable $X$ with $m$ outcomes $\{x_1,\ldots,x_m\}$, the entropy, a measure of uncertainty and denoted by $H(X)$, is defined as

$$H(X) = -\sum_{i=1}^{m} p(x_i) \log p(x_i), \qquad (3)$$

where $p(x_i)$ is the probability that $X$ equals to $x_i$.

Let $p^+$ and $p^-$ denote the probability of a document belonging to positive and negative category respectively, then $p^+$ and $p^-$ can be estimated as

$$p^+ \approx \frac{a/N^+}{a/N^+ + c/N^-}, p^- \approx \frac{c/N^-}{a/N^+ + c/N^-}. \qquad (4)$$

Here we divide $a$ and $c$ by $N^+$ and $N^-$ respectively. This makes our method work for unbalanced datasets. According to formula (3), if term $t_i$ occurs in a document, the degree of uncertainty of this document belonging to a category is

$$h = -p^+ \log_2 p^+ - p^- \log_2 p^-. \qquad (5)$$

Obviously, if the documents containing term $t_i$ distribute evenly over different categories, the entropy $h$ will be large. In contrast, if the documents containing term $t_i$ distribute unevenly over different categories, the entropy $h$ will be relatively small. However, we hope that the more uneven the distribution of documents where term $t_i$ occurs, the larger the weight of $t_i$ is. And that the entropy $h$ is between 0 and 1, so we define the original formula of term weight as

$$g_i = 1 - h. \qquad (6)$$

We call the scheme formulated by the (6) as nature entropy (*ne*). It seems that *ne* can be used as the weights of terms directly and will perform well. Unfortunately, *ne* may severely suffer from the problem of over-weighting. To avoid over-weighting, we modify *ne* with two regularized techniques: (1) using add-one smoothing for all terms and (2) adding a bias term to shrink the ratios between term weights. Regularized version of *ne* is formulated as

$$g_i = b_0 + (1-b_0)(1-h).  \quad (7)$$

Here $b_0 \in [0, 1]$ is the bias term, whose value controls the trade-off between over-weighting and under-weighting. We name the proposed scheme formulated by (7) regularized entropy (*re*).

## 4 Datasets and Experimental setup

### 4.1 Datasets

We conduct experiments on both topical and sentiment classification datasets. Detailed statistics are shown in table 4. The sentiment classification datasets are:

**RT-2k:** de facto bechmark for sentiment analysis, containing 2000 movie reviews (Pang and Lee, 2004). The reviews are balanced across sentiment polarities.

**IMDB:** a large movie review dataset containing 50k reviews (25k training, 25k test), collected from Internet Movie Database (Mass et al., 2011). The reviews are also balanced across sentiment polarities.

The topical classification datasets are:

**N-WiX, N-GrX, N-MaI, N-MoA, N-PoR:** The Newsgroups dataset with headers removed. Classification task is to classify which topic a document belongs to. N-WiX: comp.os.ms-windows.misc vs. comp.windows.x, N-Grx: comp.graphics vs. comp.windows.x, N-MaI: comp.sys.ibm.pc.hardware vs. comp.sys.mac.hardware, N-MoA: rec.motorcycles vs. rec.autos, N-PoR: talk.politics.misc vs. talk.religion.misc. The documents are approximately balanced across topics.

**R-Ear, R-Acq, R-Mon, R-Gra, R-Cru:** Reuters-21578 dataset, using documents from top 5 largest categories. The classification task is to retrieve documents belonging to a given topic. R-Ear: earn, R-Acq: acq, R-Mon: money-fx, R-Gra: grain, R-Cru: crude. The highly imbalanced category distribution of Reuters-21578 makes it significant among other datasets.

| Dataset | RT-2k | IMDB | N-WiX | N-GrX | N-MaI | N-MoA | N-PoR | R-Ear | R-Acq | R-Mon | R-Gra | R-Cru |
|---|---|---|---|---|---|---|---|---|---|---|---|---|
| $L$ | 762 | 271 | 419 | 419 | 419 | 419 | 419 | 130 | 130 | 130 | 130 | 130 |
| $N^+$ | 1000 | 25k | 985 | 973 | 982 | 996 | 775 | 3753 | 2131 | 601 | 528 | 510 |
| $N^-$ | 1000 | 25k | 988 | 988 | 963 | 990 | 628 | 3770 | 5392 | 6922 | 6995 | 7013 |

Table 4. Dataset statistics. $L$: average number of unigram tokens per document. $N^+$: number of positive examples. $N^-$: number of negative examples.

### 4.2 Experimental setup

We lower-case all words but do not perform any stemming or lemmatization. We restrict the vocabulary to all tokens that occurred at least 3 times in the training set. Unless otherwise specified, only unigrams are used as feature terms. Support vector machine (SVM) is used as the classifier in our experiments. Specially, we adopt the L2-regularized L2-loss linear SVM and the implementation software is LIBLINEAR (Fan et al., 2008). Except setting $C = 0.3$ for SVM on IMDB, $C = 1.0$ is used for other datasets. Tuning of bias $b_0$ is done by testing the models on the held-out portion of the training data, and then the models were re-trained with the chosen $b_0$ using the entire training data.

For RT-2k, there are no separate test sets and hence we use 10-fold cross validation. The overall result is the average accuracy across 10 folds. The standard train-test splits are used on IMDB and Newsgroups datasets. For Reuters-21578, training and test documents are separated according to ModApte split.

Scheme *tp* is used as local term weighting method for RT-2k and IMDB due to its superior performance on the two sentiment datasets. For newsgroups datasets, *atf* is adopted. For Reuters-21578 datasets, we choose *tf* as the local term weighting scheme. Cosine normalization is used for all term weighting schemes in our experiments.

For Reuters-21578 dataset, F1-score is used as the evaluation metric due to the highly imbalanced category distribution. Classification accuracy is adopted for other datasets.

| Dataset | no | idf | ig | chi | mi | mi' | rf | dsidf | dsbidf | ne | re |
|---|---|---|---|---|---|---|---|---|---|---|---|
| RT-2k (1) | 87.20 | 87.65 | 86.90 | 86.60 | 88.05 | 88.05 | 86.95 | 88.25 | 88.40 | 80.80 | **89.30** |
| RT-2k (2) | 87.80 | 89.20 | 88.00 | 87.55 | 89.65 | 89.65 | 87.45 | 89.75 | 89.95 | 85.55 | **90.30** |
| IMDB (1) | 88.76 | 89.20 | 86.73 | 86.65 | 89.41 | 89.41 | 88.60 | 89.26 | 89.36 | 86.87 | **89.68** |
| IMDB (2) | 90.11 | 91.24 | 88.04 | 87.92 | 91.60 | 91.60 | 90.16 | 91.76 | 91.75 | 90.38 | **91.86** |
| N-WiX | 89.73 | 92.90 | 90.11 | 88.97 | 92.27 | 92.40 | 88.85 | 92.52 | 91.76 | **93.66** | 92.78 |
| N-GrX | 86.86 | 88.39 | 87.12 | 87.12 | 90.43 | **90.56** | 87.88 | 89.80 | 89.67 | 88.90 | 89.54 |
| N-MaI | 88.41 | 90.09 | 90.60 | 90.09 | 94.08 | 94.08 | 86.87 | 94.72 | 94.85 | 93.82 | **94.98** |
| N-MoA | 96.10 | 96.98 | 93.83 | 93.45 | 97.36 | 97.36 | 94.46 | 97.10 | 97.10 | **98.11** | 97.61 |
| N-PoR | 86.81 | 88.59 | 87.52 | 86.98 | 90.02 | 90.02 | 86.63 | 90.20 | 90.20 | **90.73** | 89.66 |
| R-Ear | 98.61 | 98.61 | 96.31 | 96.40 | 98.66 | 98.66 | 98.47 | 98.56 | 98.41 | 98.61 | **99.07** |
| R-Acq | 96.07 | 96.84 | 91.07 | 91.65 | 95.43 | 96.47 | 95.67 | 96.86 | 95.58 | 96.39 | **96.93** |
| R-Mon | 98.22 | 98.57 | 93.48 | 94.37 | 96.55 | 98.94 | 97.86 | 98.95 | 97.10 | **99.29** | 98.94 |
| R-Gra | 93.28 | 96.55 | 94.62 | 95.88 | 94.66 | 96.55 | 95.42 | 96.21 | 94.98 | 97.36 | **97.76** |
| R-Cru | 90.61 | 92.11 | 83.01 | 88.89 | 91.36 | 92.59 | 93.17 | 92.68 | 88.67 | 92.31 | **93.62** |

Table 5. Results of our *re* (and *ne*) against existing term weighting schemes. The best results are in bold. For RT-2k and IMDB, (1) indicates using unigrams as feature terms, and (2) indicates using both unigrams and bigrams. For other datasets, only unigrams are used as feature terms. *mi'* (see formula (8) in section 5.1) is our improved version of *mi*. For comparison, we also include a special scheme *no*, i.e., no global term weighting scheme is used.

## 5 Experimental Results

Code to reproduce our experimental results is available at https://github.com/hymanng.

### 5.1 Comparison with existing schemes

We compare our *re* and *ne* against various existing global term weighting schemes. Results are shown in table 5. We summarize the results as follows.

(1) Our *re* is a robust performer. It performs well on all datasets, including sentiment and topical, balanced and imbalanced datasets. Specially, it achieves the best results on 9 of 14 tasks. The performance of *ne* is, however, very different across datasets. On RT-2k and IMDB, it provides much lower accuracy than *re*. This indicates *ne* suffers from over-weighting, and *re* benefits significantly from regularization techniques (*re* is the regularized version of *ne*). On newsgroups datasets, *ne* performs closely or even better than *re*, indicating that over-weighting is not serious for newsgroups datasets.

(2) *mi* performs well on balanced datasets, but poorly on the imbalanced ones: R-Acq, R-Mon, R-Gra and R-Cru. To improve *mi*, we divide $a$ and $c$ by $N^+$ and $N^-$ respectively, and let $N/N^+$ and $N/N^-$ be 2, getting *mi'*:

$$mi' = \log_2 \max(\frac{2a/N^+}{a/N^+ + c/N^-}, \frac{2c/N^-}{a/N^+ + c/N^-}). \quad (8)$$

As seen in table 5, *mi'* performs well on all datasets, including balanced and imbalanced ones. It produces the same results with *mi* on balanced datasets, but much better results on imbalanced datasets.

(3) Other results. *ig* and *chi* perform very poorly. *idf* performs remarkably well on all datasets, including sentiment and topical, balanced and imbalanced datasets, despite not the best scheme. *dsidf* is an excellent performer, better than *idf*. *dsbidf* performs well on balanced datasets, but provides very poor results on imbalanced ones. *rf* does not yields good results, even underperforms *no*. This is not surprising due to its discrimination against the terms that appear more frequently in the negative category.

To further illustrate the robustness of *re*, we reduced the dataset size on RT-2k, IMDB, N-MaI and R-Cru, and compare *re* with *no*, *idf*, *ig*, *mi'* and *dsidf*. Results are shown in figure 3. Overall, *re* performs best with different number of examples. *ig* still provides very poor results. Note that *dsidf* and *mi'* even underperforms *no* when the number of examples are less than 1600 on RT-2k. The possible reason is that they suffer from over-weighting with the reduced data size. One observation to support this is that *re* needs larger $b_0$ with smaller data size.

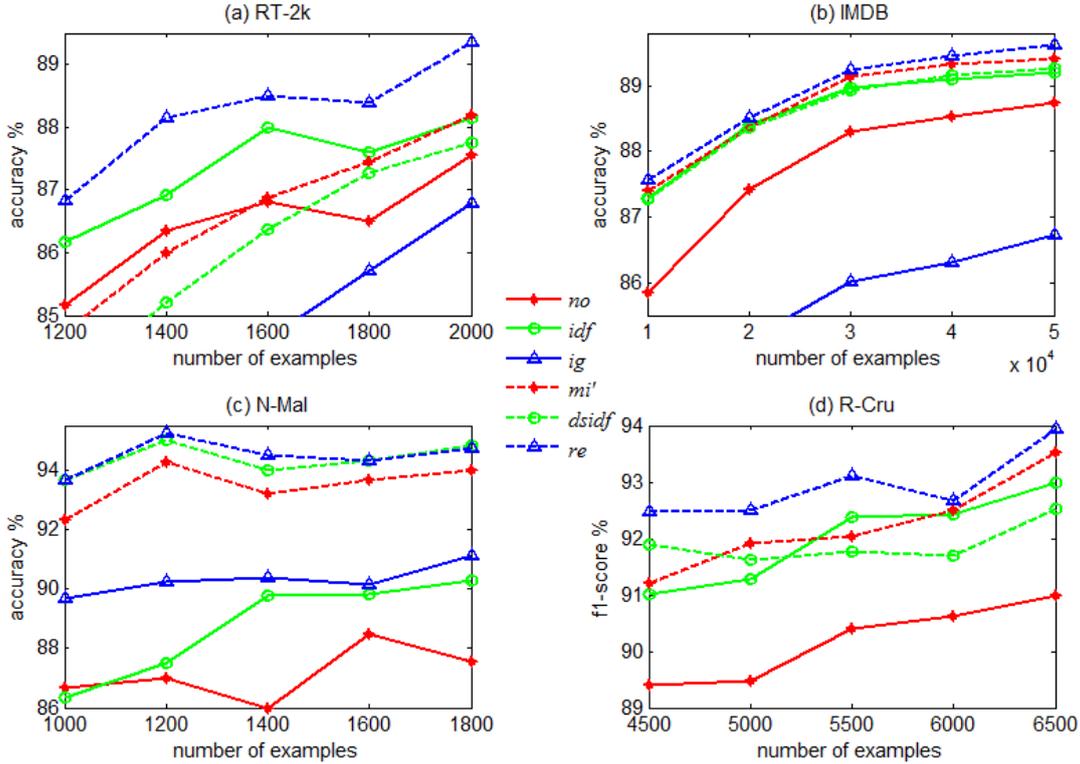

Figure 3. Results of *re* against existing schemes with different number of examples on RT-2k, IMDB, N-MaI and R-Cru. Examples are randomly selected to ensure the same category distributions with the whole datasets. For R-Cru, *ig* only gets f1-score of about 83%, far lower than results of other schemes, so its performance is not illustrated.

### 5.2 Comparison of scaling functions

As scaling function changes the ratio between term weights, unsuitable choices of it could result in over-weighting or under-weighting problem. To confirm this, we compare the performance of scaling functions *f1-f7* against *f0*. Results are shown in table 6. The first thing to note is that *f1* performs poorly on all datasets, especially for RT-2k, IMDB, N-PoR, and Reuters-21578 datasets. This is not surprising as *f1* seriously amplifies ratio between term weights, leading to severe over-weighting problem. *f7* also performs poorly, but for the opposite reason. It shrinks too much the ratio between term weights, resulting in under-weighting problem. *f2-f6* shrinks the ratio between term weights to more reasonable range, and thus provides better results than *f1* and *f7*. Specially, *f5* performs well on all datasets, and achieves the best results on 6 of 14 tasks.

### 5.3 Effect of bias term

As analysed in section 4, the value of $b_0$ controls the trade-off between over-weighting and under-weighting. If $b_0$ is too small, over-weighting occurs and harms the performance. If $b_0$ is too large, under-weighting occurs. To confirm the effect of bias term $b_0$, we test the performance of *re* with different $b_0$ value. Figure 4 presents the results on RT-2k, IMDB, N-MaI and R-Cru. Overall, the performance of *re* and the value of $b_0$ exhibits an inverted U-shaped relationship. Without bias term (i.e. $b_0 = 0$), *re* does not performs well, even underperforming *idf* on RT-2k and IMDB. This is due to over-weighting problem with too small $b_0$. If $b_0$ is too large, *re* still performs poorly. This is due to under-weighting problem with too large value of $b_0$. *re* beats *idf* with a wide range of $b_0$ settings, indicating *re* is a robust performer. Note that for N-MaI, *re* provides the best results with very small $b_0$ value, increasing $b_0$ harms the performance. This indicates that over-weighting is not a serious problem for this dataset.

| Dataset | f0 | f1 | f2 | f3 | f4 | f5 | f6 | f7 |
|---|---|---|---|---|---|---|---|---|
| RT-2k (1) | 88.10 | 81.45 | **89.05** | 88.45 | 87.90 | **89.05** | 88.85 | 88.00 |
| RT-2k (2) | 89.80 | 82.70 | 89.80 | 89.25 | 89.75 | **90.25** | 90.05 | 88.60 |
| IMDB (1) | 89.39 | 86.60 | 89.59 | 89.30 | 89.26 | **89.62** | 89.61 | 89.04 |
| IMDB (2) | 91.54 | 88.45 | 91.28 | 91.04 | 91.76 | 91.72 | **91.81** | 90.59 |
| N-WiX | **93.28** | 91.13 | 92.14 | 91.00 | 92.52 | 92.14 | 92.77 | 91.38 |
| N-GrX | 88.64 | 88.14 | 89.03 | 89.15 | **89.80** | 88.90 | 88.64 | 88.14 |
| N-MaI | 93.95 | 93.44 | 93.69 | 92.66 | **94.72** | 94.21 | 94.21 | 91.25 |
| N-MoA | 96.72 | 95.97 | 96.85 | 96.22 | **97.10** | **97.10** | 96.98 | 96.35 |
| N-PoR | 88.59 | 84.85 | **90.55** | 88.95 | 90.20 | 89.48 | 89.48 | 88.24 |
| R-Ear | 98.13 | 97.19 | 98.56 | **98.66** | 98.56 | 98.56 | 98.47 | 98.56 |
| R-Acq | 96.86 | 95.47 | 96.39 | 96.31 | 96.85 | 96.55 | **96.93** | 95.92 |
| R-Mon | 96.19 | 85.71 | 98.60 | 98.23 | 98.95 | **99.30** | 98.95 | 98.23 |
| R-Gra | 96.29 | 92.75 | 96.24 | 95.79 | 96.21 | **96.95** | 96.24 | 96.15 |
| R-Cru | 90.48 | 85.55 | 92.68 | **93.50** | 92.68 | 93.01 | 92.73 | 91.77 |

Table 6. Results of different scaling functions. The best results are in bold. For RT-2k and IMDB, (1) indicates using unigrams as feature terms, and (2) indicates using both unigrams and bigrams. For other datasets, only unigrams are used as feature terms.

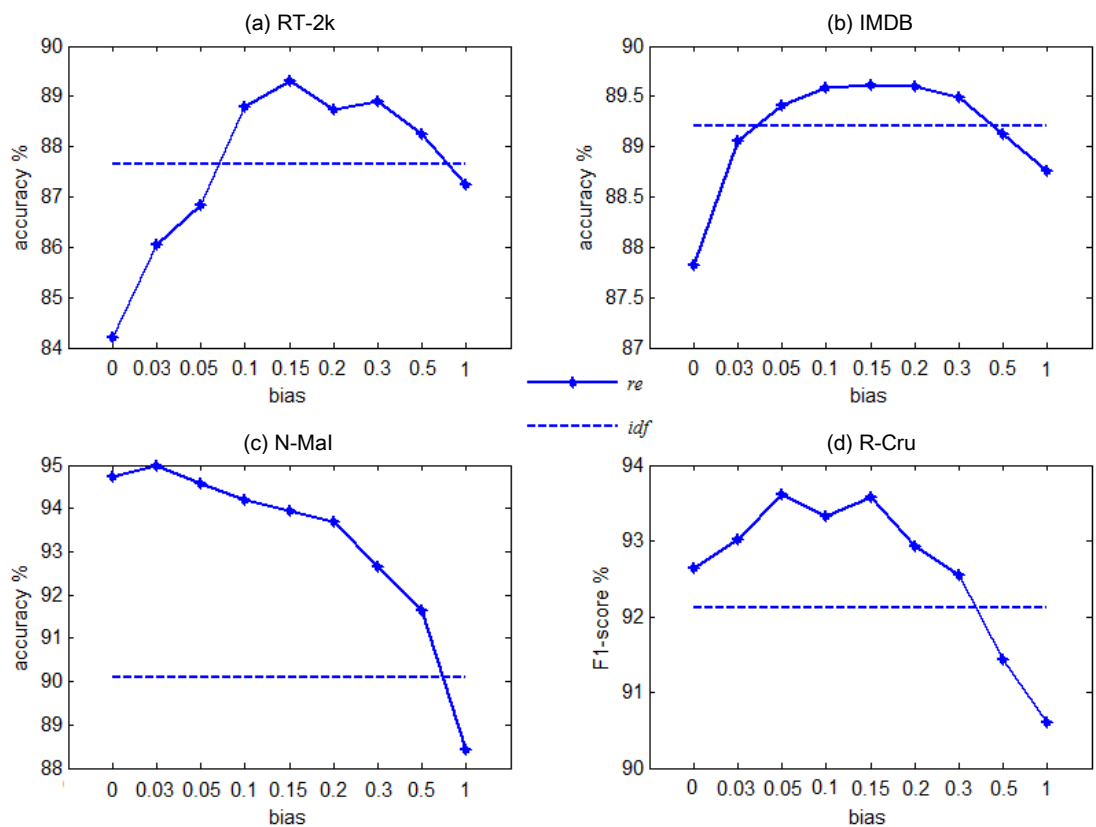

Figure 4. Result of our *re* with different $b_0$ values on RT-2k, IMDB, N-MaI and R-Cru. For comparison, the results of *idf* are also included.

## 6   Conclusions

In this paper we have presented the importance of balancing between over-weighting and under-weighting in supervised term weighting for text categorization. Over-weighting is a new concept proposed in this paper. It is caused by the improper handling of singular terms and the unreasonably too large ratios between weights of different terms. To reduce over-weighting, three regularization techniques, namely add-one smoothing, sublinear scaling and bias term, are introduced. Add-one smoothing can be used for singular terms to avoid over-weighting. Sublinear scaling and bias term shrink the ratios between weights of different terms. If the scaling functions scale down the weights too much or bias term is too large, ratios between term weights become too small. Hence under-weighting occurs. Experiments on both topical and sentiment classification datasets have shown that regularization techniques could significantly enhance the performance of supervised term weighting.

More advanced, a new supervised term weighting scheme, *re*, is proposed under the insight of balancing between over-weighting and under-weighting. *re* bases on entropy, which is used to measure a term's distribution across different categories. The value of bias term in *re* controls the trade-off between over-weighting and under-weighting. Empirical evaluations show that *re* performs well on various datasets, including topical and sentiment, balanced and imbalanced ones.

**Acknowledgements**

This work was supported in part by National Natural Science Foundation of China under grant 61371148.